\documentstyle[12pt]{article}

\title{\bf{Response to Fackerell's Article}}

\author{C. O. Alley\footnote{Department of Physics, University of
Maryland, College Park, Maryland 20742. e-mail: coalley@physics.umd.edu.}
\ and H. Y{\i}lmaz\footnote{Hamamatsu Photonics, K.K., Hamamatsu City,
430-8587 Japan and Electro-Optics
Technology Center, Tufts University, Medford, Massachusetts 20155.}}
\date{15 August 2000}
\begin{document}
\maketitle

\begin{abstract}
   E. D. Fackerell claims:  1) that Alley and Y{\i}lmaz treatment
of parallel slabs in general relativity is wrong because the
Y{\i}lmaz metric used is not a solution of the field equations
of general relativity; 2) he also claims that the correct
treatment of the parallel slab problem in general relativity
must be based on the so-called Taub metric. We show below that
both of Fackerell's claims are false.  His first claim is
based on his failure to distinguish the matter-free regions
and the regions with matter.  His second claim is based on his
failure to recognize that for the Taub metric the left-hand
side, hence also the right-hand side of the field equations, 
are identically zero everywhere. Thus no material systems can
be treated via the Taub metric.
\end{abstract}
\section{Introduction}
In an article titled ``Remarks on the Y{\i}lmaz and Alley Papers" E.D.
Fackerell  \cite{1} claims that the Y{\i}lmaz metric, used by H.
Y{\i}lmaz \cite{2} and
C.O. Alley \cite{3} for the treatment of parallel slabs in general
relativity,
is not a solution to the field equations of general relativity.  His train
of
thought in making this statement seems to be as follows:

The Y{\i}lmaz metric used in the said treatments is of the form
\begin{equation}
ds^2 = e^{\textstyle-2\phi} dt^2 - e^{\textstyle2\phi} [e^{\textstyle 2
\epsilon
\phi} (dx^2 + dy^2) + e^{\textstyle 4\epsilon \phi} dz^2]
\end{equation}
\begin{equation}
\phi = az + {\textstyle\frac{1}{2}} \sigma z^2 + C
\end{equation}
where $\epsilon = \pm 1$ for general relativity.  [$\epsilon = 0$ for the
Y{\i}lmaz theory.]  But Fackerell, based on some calculations of his, seems to
conclude that the field equations will everywhere require $-g_{zz}$ to be
\begin{equation}
-g_{zz} = e^{\textstyle2(1 + 2\epsilon)\phi +2\ln |\partial_z \phi| + K}.
\end{equation}

Therefore, he says, the Y{\i}lmaz metric will be valid if and only if
$\partial_z\phi =$ {\it Constant} so that one can set $2\ln
|\partial_z\phi | + K = 0$ to regain the Y{\i}lmaz form.  Since $\partial_z\phi$ is
not a constant in the Y{\i}lmaz field $\phi$ in Eq. (2) above, the metric used in
the treatment, he claims, cannot be a solution to the field equations of general
relativity! \\

However, a simple calculation, which can be done by hand, shows that the Y{\i}lmaz
metrics for $\epsilon = \pm 1$ satisfy the field equations of general relativity
everywhere exactly.  If Fackerell did this calculation he would have seen that it
is indeed true.  Therefore the question is not whether Y{\i}lmaz and Alley used
wrong metrics in their works but rather how and where precisely Fackerell himself
made a fundamental error in his calculations.\\

\section{Fackerell's Fundamental Error}

To find this out, we shall first verify that Y{\i}lmaz' metrics are indeed
solutions of the required field equations.  Fackerell's error will be clearly seen
in the course of this exercise.  The general relativity equations to satisfy with
$\tau_{\mu}^{\nu}$ as a matter tensor are 
\begin{equation}
{\textstyle\frac{1}{2}} G_{\mu}^{\nu} = \tau_{\mu}^{\nu}.
\end{equation}

There are two distinct cases with two sub-cases each: [Note 1]
\newpage
\[
{\rm A):} \ \sigma = 0 , \ \epsilon = \pm 1 \ \ ({\rm Matter-free \
regions})
\]
\[
{\rm B):} \ \sigma \neq 0 , \ \epsilon = \pm 1 \ \ ({\rm
Regions \ with \ matter})
\]

where $\sigma$ corresponds to the ordinary Laplacian of $\phi$.  The Y{\i}lmaz
metric is of the form

\begin{equation}
ds^2 = e^{\textstyle-2\phi} dt^2 - e^{\textstyle \mu} (dx^2 + dy^2) -
e^{\textstyle\eta} dz^2.
\end{equation}

where $\phi$, $\mu$ and $\eta$ are functions only of the coordinate z.  Using the
shorthand $(^\prime) = \partial_z\phi$, the general relativity equations (4) for
metric (5), after multiplying by $8e^{\eta}$, are:
\begin{equation}
 \mu^{\prime} (3\mu^{\prime} - 2\eta^{\prime}) +
4\mu^{\prime \prime}= 8e^{\eta}\tau_0^0
\end{equation}
\begin{equation}
4\phi^{\prime \, 2} + (\mu^{\prime} -
2\phi^{\prime}) (\mu^{\prime} - \eta^{\prime}) + 2(\mu^{\prime
\prime} -2\phi^{\prime \prime})= 8e^{\eta}\tau_1^1 =
8e^{\eta}\tau_2^2
\end{equation}
\begin{equation}
\mu^{\prime}(\mu^{\prime} - 4\phi^{\prime}) = 8e^{\eta}\tau_3^3.
\end{equation}

Let us first consider the matter-free case (A): \\ For the Yilmaz metric, where
$\mu = 2(1 +
\epsilon) \phi$ and $\eta = 2(1 + 2\epsilon ) \phi$, all terms quadratic in
$\phi^{\prime}$ cancel. (The algebra is exhibited in [Note 2].) The equations (6)
and (7) in this case become: 
\begin{equation}
 (1 + \epsilon ) \phi^{\prime \prime} = 0 
\end{equation} 
\begin{equation} 
{\textstyle\frac{1}{2}}\epsilon
\phi^{\prime
\prime} = 0 \end{equation}
with the solution
\begin{equation}
\phi^{\prime \prime} = 0
\end{equation}
\begin{equation}
\phi = az + C
\end{equation}
where $a$ and $C$ are constants, hence $\phi^{\prime} =${\it Constant} is
automatic.  Setting $2\ln |a| + K = 0$ in equation (3) above, as Fackerell himself
suggests, we have
\begin{equation}
-g_{zz} = e^{\textstyle2(1 + 2\epsilon ) \phi}
\end{equation}
as originally stated by Y{\i}lmaz. \\ \\ \indent Let us now consider also the
regions with matter, case (B): \\ The $\tau^{\nu}_{\mu} \not= 0$
equations (6) - (8) are satisfied with the matter tensor of the form
\begin{equation}
\tau^{\nu}_{\mu} = e^{-\eta}\rm diag { [(1 +
\epsilon, \ \epsilon /2, \ \epsilon /2, \ 0)\sigma]}
\end{equation}
where by directly substituting $\mu = 2(1 + \epsilon ) \phi$, $\eta = 2(1 +
2\epsilon ) \phi$, all terms quadratic in $\phi^{\prime}$ again cancel for
$\epsilon = \pm 1$ (as shown in [Note 2]) and we have
\begin{equation}
(1 + \epsilon) \phi^{\prime \prime} = (1 + \epsilon) \sigma 
\end{equation}
\begin{equation}
{\textstyle\frac{1}{2}} \epsilon
\phi^{\prime \prime} = {\textstyle\frac{1}{2}} \epsilon \sigma
\end{equation}
with the solution
\begin{equation}
\phi^{\prime \prime} = \sigma
\end{equation}
\begin{equation}
\phi = az + \frac{1}{2} \sigma z^2 + C
\end{equation}
as stated by Alley and Y{\i}lmaz.  But now there are no logarithmic terms in
the
exponential because $\phi^{\prime \prime}$ terms causing them are taken
away by equating them to 
the matter terms.  The result is again the form of equation (13) above
\[
-g_{zz} = e^{\textstyle2(1 + 2\epsilon )\phi}
\]
as originally stated by Y{\i}lmaz.\\

The whole discussion can be reduced into the realization that
\begin{equation}
\phi^{\prime \prime} = \left\{ \begin{array}{cc} 0 & \ \ \ ({\rm 
Outside \ a \ slab}), \\
\sigma & ({\rm Inside \ a \ slab})\end{array} \right.
\end{equation}
Nothing else is needed, not even Fackerell's way of removing the
logarithmic
terms above, since the form given by Y{\i}lmaz makes it
superfluous. [Note 3]\\

Thus, whether we like it or not, the Y{\i}lmaz metric is a solution of the
field equations of general relativity, both in the matter-free regions and in
the regions with matter.  Then, where did Fackerell make his mistake?  In the
light
of the above discussion we do not have far to look.  Since in a
matter-free region
$\phi$ is linear $(\phi^{\prime\prime} = 0$) Fackerell thought
this would contradict the quadratic term in equation (18) above 
\[
\phi = az + {\textstyle\frac{1}{2}} \sigma z^2 + C.
\]
But those are different regions with different material
contents as
indicated above.  Linearity in coordinates of a potential in one region does not
prevent it
from being nonlinear in coordinates in a different region.  Thus, contrary to
this fact, Fackerell is requiring that the linearity of $\phi$ in
the matter-free region be
imposed on the region with matter.  This is equivalent to
requiring that there
be no matter terms at all, anywhere.  A more elementary calculational error is
hardly imaginable.\\

We should really stop here and go no further with Fackerell's paper.  All
his
other asides such as his appeals to authority and his prima facie case,
etc.,
are irrelevant    since his main tenet is false. Therefore it is
useless to go over every conceivable argument. We must however,
mention just one additional error, because this one is so
hopelessly hidden that the reader, not being able to penetrate,
might think he has something. \\

\section{Fackerell's Second Error: His Treatment of the Parallel Slab
Problem in General Relativity }

Let us now also examine Fackerell's own treatment of the parallel slabs. 
Fackerell claims that  the appropriate metric to use for the
\underline{slab} calculations
is the Taub metric
\begin{equation}
ds^2 = f^{-1/2} dt^2 - f (dx^2 + dy^2) - f^{-1/2} dz^2
\end{equation}
where
\begin{equation}
f = (1 + kz).
\end{equation}
However, he does not use the slabs as originally defined, and he does not use
equations of motion.  Instead, he considers infinitely thin sheets.  Under
some sophisticated looking maneuvers he claims he gets certain results which
sound impressive.  But it can be shown by a simple calculation that, for the Taub
metric the left-hand side (hence also the right-hand side) of the field
equations are identically zero.  Again, if Fackerell performed this
calculation he would have seen that it is true.  In any case the proof is so simple
that we shall here reproduce it: \ First, identify from the Taub metric by comparing
with the general metric, equation (5), that $\mu = 4\phi$, $\eta = -2\phi$.  Then
simply calculate to get,
\begin{equation}
e^{4\phi} = f= 1 + kz
\end{equation}
\begin{equation}
\phi = {\textstyle\frac{1}{4}} \ln (1 + kz)
\end{equation}
\begin{equation}
\phi^{\prime} = (k/4)/(1 + kz) 
\end{equation}
\begin{equation}
\phi^{\prime \prime} = -(k^2/4)/(1 +
kz)^2 = -4\phi^{\prime \, 2}.
\end{equation}
Finally, substituting into the required field equations (6) - (8)  one finds that
all expressions on the left-hand side are identically zero.  (The steps in
this calculation are exhibited in [Note 4].) Therefore the right-hand side is
\begin{equation}
\tau^{\nu}_{\mu} = \ {\rm diag} \ [(0,0,0,0)].
\end{equation} 
This is because, unlike the Yilmaz metrics, the term linear in $\phi^{\prime
\prime}$ (that is, the Laplacian) also disappears and nothing is left to represent
matter. \\
 
       As to the equations of motion, we shall then have
\begin{equation}
\sigma d^2 x_{\mu}/ds^2 ={\textstyle\frac{1}{2}} \partial_{\mu} g_{\alpha \beta}
\tau^{\alpha \beta} = 0.
\end{equation}
This is clearly the answer to the problem of motion with the Taub metric. But as we
have just said, Fackerell does not use the equations of motion of the theory,
and he does not use the definition of the slab as originally intended. 
Instead, he introduces infinitely thin sheets.  This is presumably to get
around the difficulty that he cannot introduce a matter density
$\tau_{\mu}^{\nu}$.  The truth is that as long as he uses the Taub metric he cannot
introduce any matter at all.  Under these circumstances how can the ``junction
conditions'' etc., be of any help?  Can there be a magic in the junction conditions
so as to create something out of absolutely nothing? [Note 5]\\

Y{\i}lmaz and Alley do not use such devices.  When they say slab, they mean slab,
when they say equations of motion, they mean equations of motion.  They do
not use junction conditions to supplant the equations of motion nor do they
stray from the boundary conditions of the standard field theory.  As Fackerell
himself points out, the ``junction conditions'' he uses are not necessarily
equivalent to the boundary conditions of the standard field theory.  Fackerell's
deductions from the Taub metric are clearly false.\\

\section{Discussion}

    We have been informed that during a recent discussion on the internet
newsgroup $\it{sci.physics.relativity}$, a respected member of the group
stated: \\ \\
 ``Alley and Y{\i}lmaz's claims along these lines
have been demolished, thoroughly and in
detail, in a paper by Edward Fackerell,
...[ref.\,[1] of this paper]..., Fackerell
shows that Y{\i}lmaz and Alley made a careless
mistake in their calculation, and he gives the
correct calculation, for both flat plates and
spherical shells.''\\

 We cannot expect everyone to dig
into the truth of the matter but we must urge Fackerell  to honorably
retract
his position.  This sort of thing is not doing any service to science --
nor
any justice to the scientific truth. \\ \\

\noindent \Large \bf Acknowledgment \rm \normalsize \\ \\
We wish to thank Kirk Burrows for carrying out the detailed calculations
associated with our treatment of the two parallel slabs problem \cite{4}. His
expert knowledge of the symbolic calculational capabilities of \it Mathematica
\rm has been invaluable in our continuing investigations of the theories of
Y{\i}lmaz and of Einstein.

\noindent  \Large\bf Notes \rm \normalsize \\

 [Note 1] Fackerell treats the cases $\epsilon = + 1$
and
$\epsilon = -1$  separately from the start.  This gets
confusing and repetitive.  We treat
them uniformly with $\epsilon$ as a parameter and only at the end set
$\epsilon =
+1$ and $\epsilon = -1$. \\

 [Note 2]   Substituting the expressions $\mu = 2(1 +
\epsilon) \phi$ and $\eta = 2(1 + 2\epsilon ) \phi$ into the left-hand sides of
equations (6) - (8), one obtains the following: \\ 

\noindent Left-hand side of Equation (6):
\[
\mu^{\prime} (3\mu^{\prime} - 2\eta^{\prime}) +
4\mu^{\prime \prime},
\]
\[
2(1 + \epsilon) \phi^{\prime} [6(1+\epsilon)\phi^{\prime} -
4(1+2\epsilon)\phi^{\prime}]+4[2(1+\epsilon)\phi^{\prime \prime}], 
\]
\[
2(1 + \epsilon) \phi^{\prime} [2(1 -
\epsilon)\phi^{\prime}]+8(1+\epsilon)\phi^{\prime \prime},
\]
\[
 4\phi^{\prime\, 2}(1-\epsilon^2)+8(1+\epsilon)\phi^{\prime \prime}
\] 
Left-hand side of Equation (7):

\[
4\phi^{\prime \, 2} + (\mu^{\prime} -
2\phi^{\prime}) (\mu^{\prime} - \eta^{\prime}) + 2(\mu^{\prime
\prime} -2\phi^{\prime \prime}),
 \]
\[ 
4\phi^{\prime \, 2} + [2(1 + \epsilon) \phi^{\prime} - 2\phi^{\prime}][2(1 +
\epsilon) \phi^{\prime} - 2(1 + 2\epsilon) \phi^{\prime}] +
2[2(1+\epsilon)\phi^{\prime \prime} - 2\phi^{\prime \prime}],
\]
\[
4\phi^{\prime \, 2} + [2\epsilon\phi^{\prime}][-2\epsilon\phi^{\prime}] +
4\phi^{\prime \prime},
\]
\[
4\phi^{\prime\ 2}(1-\epsilon^2) + 4\epsilon\phi^{\prime \prime} 
\]
Left-hand side of Equation (8):
\[
 \mu^{\prime}(\mu^{\prime} - 4\phi^{\prime}), 
\]
\[
2(1+\epsilon)\phi^{\prime} [2(1+\epsilon)\phi^{\prime} - 4\phi^{\prime}], 
\]
\[
2(1+\epsilon)\phi^{\prime}[-2\phi^{\prime} + 2\epsilon\phi^{\prime}],
\]
\[
-4\phi^{\prime\ 2}(1-\epsilon^2).
\]
The factor $(1-\epsilon^2)$ appears in the left-hand side of each equation
multiplying $\phi^{\prime \, 2}$. To satisfy Equation (4), one must have 
$(1-\epsilon^2) = 0$, leading to the values
$\epsilon = \pm 1$ for the general relativity solutions. Equations (9) and (10) in
the text are obtained after dividing the left-hand sides above by 8.\\

 [Note 3] The only possible demand Fackerell can have would be, that
the matter tensor should be
 $\tau^{\nu}_{\mu} = e^{-\eta}{\rm diag} \ [(\sigma
,0,0,0)]$ for Newtonian correspondence.  Ah $\cdots$, but that requires $\epsilon
= 0$, which is precisely the Y{\i}lmaz Theory!  Some general relativists refer to
the matter tensor
$\tau^{\nu}_{\mu} = e^{-\eta} \rm diag { [(1 +
\epsilon, \ \epsilon /2, \ \epsilon /2, \ 0)\sigma]}$, with $\epsilon = \pm 1$,
shown here to be a solution of the general relativity field equations for slabs, as
``peculiar matter.''  However, the standard non-isotropic Schwarzschild metric
solutions of general relativity have matter tensors of exactly this form! \cite{3}
\\
 
 [Note 4]  To see that the Taub metric is
devoid of physical content, substitute into the left-hand sides of the field
equations (6) - (8)  the relations which hold in the Taub metric: $\mu = 4\phi$,
$\eta = -2\phi$,
$\phi^{\prime
\prime} = -4\phi^{\prime \, 2}$, to obtain:
\\
\\ Left-hand side of Equation (6):
\[ 
 \mu^{\prime} (3\mu^{\prime} - 2\eta^{\prime}) +
4\mu^{\prime \prime},
 \] 
\[
 4\phi^{\prime} (12\phi^{\prime} + 4\phi^{\prime}) +
16\phi^{\prime \prime}, \] 
\[
 64\phi^{\prime\ 2} - 64\phi^{\prime\ 2} \equiv 0.
\]  
Left-hand side of Equation (7): 
\[
4\phi^{\prime \, 2} + (\mu^{\prime} -
2\phi^{\prime}) (\mu^{\prime} - \eta^{\prime}) + 2(\mu^{\prime
\prime} -2\phi^{\prime \prime}), 
\]
\[
4\phi^{\prime \, 2} + (4\phi^{\prime} - 2\phi^{\prime})(4\phi^{\prime} +
2\phi^{\prime}) + 2(4\phi^{\prime \prime} - 2\phi^{\prime \prime}),
\]
\[
 4\phi^{\prime \, 2} + 12\phi^{\prime\
2} + 4\phi^{\prime \prime}, \] \[  16\phi^{\prime \, 2} - 16\phi^{\prime \, 2}
\equiv 0.
\]
Left-hand side of Equation (8):
\[
 \mu^{\prime}(\mu^{\prime} - 4\phi^{\prime}), 
\]
\[
4\phi^{\prime}(4\phi^{\prime} - 4\phi^{\prime}) \equiv0.
\] \\

 [Note 5] The issue of interactive N-body solutions is
discussed quite generally by C.O. Alley, D. Leiter, Y. Mizobuchi and H.
Y{\i}lmaz, ``Energy Crisis in Astrophysics (Black Holes vs. N-Body Metrics),''
available at the Los Alamos e-print archive, xxx.lanl.gov, as
astro-ph/9906458 (28 June 1999).  The absence of interaction between the two
slabs is a specific example, admitting a simple analytic solution, of the 
situation in general relativity.
\\
\end{document}